\documentclass[sigconf]{acmart}
\AtBeginDocument{%
  }

\usepackage{mdframed}
\usepackage{enumitem}
\definecolor{backgroundgray}{gray}{0.9}
\newmdenv[
    backgroundcolor=backgroundgray
]{grand_challenges_box}

\usepackage{enumitem}
\usepackage[most]{tcolorbox}
\AtBeginEnvironment{tcolorbox}{\small}

\setcopyright{acmlicensed}
\copyrightyear{2026}
\acmYear{2026}
\acmDOI{XXXXXXX.XXXXXXX}
\acmConference[Conference acronym 'XX]{Make sure to enter the correct
  conference title from your rights confirmation email}{June 03--05,
  2018}{Woodstock, NY}
\acmISBN{978-1-4503-XXXX-X/2018/06}

\begin{document}


\title{A Vision for the Future of an AI-Integrated Research Ecosystem}

\author{Ryan E. Dougherty}
\email{ryan.dougherty@westpoint.edu}
\orcid{0000-0003-1739-1127}
\affiliation{%
  \institution{United States Military Academy}
  \city{West Point}
  \state{New York}
  \country{USA}
}

\author{Natalie Kiesler}
\email{natalie.kiesler@th-nuernberg.de}
\orcid{0000-0002-6843-2729}
\affiliation{%
  \institution{Nuremberg Tech}
  \city{Nuremberg}
  \country{Germany}
}

\renewcommand{\shortauthors}{Dougherty and Kiesler}

\begin{abstract}
Generative AI has infiltrated every stage of the research lifecycle: how scholarship is conducted, written, published, and reviewed.
Recent policy responses, such as ACM's authorship policy, address an immediate concern about responsible and transparent disclosure of AI use.
We argue that a focus on authorship and disclosure, although necessary, risks obscuring and ballooning a set of entrenched problems and strains within publication systems.
The central question is not about how papers and other research artifacts should incorporate AI, but how scientific communication itself should evolve when all relevant parties (authors, reviewers, readers) may rely on AI assistance.
We draw on our experience within these and other roles to illustrate two contrasting but feasible visions of 2036 with four entwined questions, namely about the purpose of papers as artifacts, reviews, human reviewers, and the incentives that bind all of them.
We argue for a shift from policing GenAI and other disruptive technologies to building the infrastructure of provenance, calibration, and accountability that would make trustworthy scholarship the default.
We conclude with three grand challenges and invite the community to a broader conversation and research pathways.
\end{abstract}



\maketitle



%

\section{Introduction}

The implications of Generative AI (GenAI) and related tools on computing education are manifold, and they have been subject to extensive research in the last couple of years~\cite{prather2023wgfullreport,raihan2025large,viane2025integrating}. 
Similarly, students, educators, and higher education institutions are adopting the technology and tools at scale~\cite{alpizar2025excited,prather2025beyond}. 
These developments helped shape the discourse on how to use GenAI to enhance learning~\cite{prather2023wgfullreport}, metacognition~\cite{prather2024thewideninggap}, or, for example, self-regulation~\cite{lopezpernas2025thedynamics}.
At the same time, we, as the computing education community, are beginning to define what an appropriate use of GenAI in computing education contexts looks like and how to communicate this through policies aligned with basic academic integrity and principles \cite{gordon2026taskforcereport,bernstein2025beyondbenefits}. 

ACM's recent authorship policy update is another example of how the computing community is moving forward and adapting to GenAI as a tool that may be used in and for research~\cite{ACM2026aipolicy}. With the new policy, ACM no longer requires disclosure of the use of AI in writing a submission and instead increases the emphasis on the authors’ responsibility for the reliability and correctness of their submissions. It specifies that ``\textit{When using Artificial Intelligence to conduct research, including the design and methodology of the research project, creation and selection of data sources, designing experiments, generation and collection of data, coding, [...] the specific use(s) of AI tools must be described in detail in the methods section of the Work.}''~\cite{ACM2026aipolicy}

The recent policy update addresses an immediate and practical concern: How to ensure the responsible and transparent use of GenAI within the current academic research and publication process. The focus on authorship and disclosure alone, however, may distract from a more fundamental question. GenAI is treated as a tool that may help with productivity or efficiency, operating in an otherwise unchanged system. It is this system that has been under increasing pressure for years, suffering from inconsistency~\cite{Lawrence2022NeurIPSExperiment}. With GenAI, we have reached a critical breaking point~\cite{zhao2026llmhallucinationswildlargescale,ACL2026HallucinatedReferences}.

In this paper, we suggest a visionary perspective. As AI increasingly participates in the scientific research, writing, evaluation, and publication of scientific knowledge, it challenges many of the assumptions upon which today's publication system was built. The central issue is therefore not how papers should incorporate AI, but how scientific communication itself should evolve in a world where authors, reviewers, and readers may use AI assistance. We are discussing these issues based on our lived experience in various roles as authors, PhD advisors, critical friends, working group leaders, reviewers, meta-reviewers, committee members, journal editors, and program chairs of the SIGCSE Technical Symposium.





    



\section{A thought experiment}

As we describe our vision of computing education research, we are rethinking today's practices surrounding conference, journal, and workshop publications in the field. Join us in the following two contrasting thought experiments. 

\textit{A bright vision.} Imagine a researcher in 2036 making a breakthrough discovery on how to significantly improve programming education. Instead of spending 9-12 months navigating publication workflows, AI systems automatically verify the human analyses, reproduce results, identify relevant prior work, and connect the new findings to the global body of scientific knowledge. AI reviews with a human-in-the-loop focus on the significance, interpretation, and plausibility of the findings. Scientific claims can quickly become public, discoverable, transparent, and continuously updated as new evidence emerges. In this future, AI does not replace human scientists and their work; it removes friction from scientific communication and processes, allowing knowledge to accumulate faster, more reliably and more transparently than ever before.

\textit{A dark future.} Now imagine a different 2036. AI systems generate manuscripts, reviews, summaries, and citations at a scale no human can comprehend or verify. Researchers no longer read papers directly but rely on AI-generated interpretations and reviews of AI-generated research. The volume of publications (along with the pressure on scientists to publish) continues to grow exponentially, while confidence in their validity and in the validity of the supposedly human reviews steadily declines. Chairs, editors, and boards keep increasing the workload of reviewers and volunteers, hoping to solve the problem. Scientific knowledge becomes abundant but increasingly opaque. In this future, the challenge is not generating discoveries but determining which discoveries can be trusted. Human researchers try not to crack under the pressure. 

\section{Questions we need to think about}
Neither of the described scenarios is intended as a prediction. Rather, they serve as a thought experiment exposing a fundamental tension. The same technologies that could make scientific knowledge more transparent could amplify existing deficits, such as the declining validity and trust in peer reviews. To structure this discussion, we focus on four interrelated questions: (1) Are paper manuscripts still an adequate artifact for sharing knowledge? (2) How should reviews evolve to foster trusted, high-quality research? (3) What should be the role of human reviewers? (4) What incentives and procedures are needed to establish recognition and ensure meaningful contributions within the scientific communities?

\subsection{Are papers the right artifact?}

A typical research paper is effectively a compressed format of work as a result of research, a linear narrative of a more or less limited length. 
The origins of research had an assumption of limited access to knowledge, and that the reader of a written manuscript would not be able to interrogate the underlying work, data, or motivation, but would rely on the authors' account. 
This assumption no longer holds today, as we now have the capabilities (i.e., infrastructure, tools, and networks) to easily share and verify not only manuscripts, but also data, software, models, and much more. This is what we currently refer to as supplementary material. Too often, such data is not recognized as an academic contribution, especially when it comes to software~\cite{kiesler2022onthelack}. Authors often do not even recognize their own research data as such, and that it is worth being published~\cite{schulz2025datadilemmaauthorsintentions}. 
Moreover, researchers are experiencing challenges and barriers when considering the publication of their research data~\cite{kiesler2024wheres}.
Most publication systems do not have a comprehensive policy in place to motivate, incentivize, and guide authors towards publishing their research data~\cite{kiesler2023why,schulz2025datadilemmaauthorsintentions}. As a consequence,
challenges to reuse, replicate or re-implement research are imminent, and persist~\cite{beardsley2018seeking}.

To foster data sharing and enhance their reuse, representatives from academia, industry, funding agencies, and scholarly publishers have developed the FAIR principles~\cite{Wilkinson2016FAIRPrinciples}. According to them, data should be \textbf{F}indable, \textbf{A}ccessible, \textbf{I}nteroperable, and \textbf{R}eusable. They emphasize the machine-readability of data through metadata, so data can be found and reused more easily -- for example, with AI.

When research data and artifacts are FAIR, the paper itself is no longer the only valid research output, but one dimension of it.
This is where provenance must appear, namely, to the living knowledge object that a paper signifies.
Through this lens, what is AI-assisted, and how, becomes visible and comprehensible.
GenAI presents a fork from which we can choose as a community. 
Either we can raise the research object as primary, or we can continue the status quo with mass-produced snapshots of research work not attached to any verifiable digital object.


\subsection{What should a review be?}

A review generally serves three main functions: verification, gatekeeping, and improvement~\cite{piratova2026curateded}.
For verification, reviews should determine whether claims are supported, references are real, and the paper carries through its logic. 
For gatekeeping, reviews make a determination of whether the paper/artifact is of sufficient quality to be considered for the venue.
And for improvement, reviews are supposed to provide advice and guidance to the authors on how to improve their paper/artifact, even if it is accepted to the venue.
Reviews are nominally an expert weighing their opinion on the value of upcoming research.

Overall, reviews do not seem to serve these three purposes well. This is supported by research on the randomness of peer review outcomes. In the context of the NeurIPS conference, for example, different committees reviewed 10\% of the submissions, revealing inconsistencies in decisions for 43 of the 166 papers, and a precision range between 0.175 and 0.495~\cite{Lawrence2022NeurIPSExperiment}. If the conference review had been run with a different committee, only half of the presented papers would have been the same. The chairs conclude that the ``conference was good for identifying poor papers, but poor for identifying good papers''~\cite{cortes2021inconsistencyconferencepeerreview}.
The analysis of the 2022 NeurIPS conference meta-reviews confirms inconsistency, bias towards irrelevant factors, miscalibration, and subjectivity as issues~\cite{goldberg2024peerreviewspeerreviews}.

Potentially AI-hallucinated references are a new challenge for reviewers. Zhao et al. analyzed 111 million references across 2.5 million papers in arXiv, bioRxiv, SSRN, and PubMed Central, finding about 147,000 hallucinated citations in 2025 alone~\cite{zhao2026llmhallucinationswildlargescale}. They conclude that ``the spread of hallucinated content has outpaced
existing safeguards'', and ``LLM hallucinations are infiltrating knowledge production at scale, threatening both the reliability and equity of future scientific discovery''~\cite{zhao2026llmhallucinationswildlargescale}. Research on the extent to which the ACM Digital Library, and, for example, recent SIGCSE Technical Symposium Proceedings, are affected is currently ongoing. 

In our experience, all of these findings are not the result of inexperienced or incompetent reviewers but are largely due to external and systemic factors.
Reviewers often remain anonymous and are not compensated for their efforts. Reviewing is neither incentivized nor recognized as an academic achievement. 
At the same time, reviews themselves are hardly ever reviewed or corrected. A careless or AI-generated review can easily slip by in large venues like conferences, which are mostly led by volunteers.
We believe that a review should be a contribution and not a verdict: published, credited, and held to the same standards as any other form of scholarship. 

In the current age of GenAI, a form of calibration seems to be the missing piece. We currently do not have a measure as to whether a review is reliable and of high quality, human or AI-assisted.
A reframing of reviews as accountable academic work with scrutiny based on provenance may help separate constructive reviews from machine-written ones. In our role as program chairs and editors, we are currently developing a process and pipeline for a tool-assisted desk review reference check. However, this can only be an initial small step.  

\subsection{Do we still need reviewers?}

Based on the previous section, we believe that asking this question is not about the level or appearance of AI within the reviewing process, but which layer of the reviewing process may be appropriate for AI use.
For example, verification of references can be tedious for humans. Software can perform such tasks well while preserving privacy with local assistants. 
Judgment, however, is a deeply human and social process. It is contestable in addition to being accountable.  
Let us remember the NeurIPS experiment, where ``the committees acted better than random, but not by a very large margin''~\cite{cortes2021inconsistencyconferencepeerreview}.

We believe that AI will not replace reviewers, but rather replace the tedious part of reviewing, forcing us to ask about the original purpose of the human within the process. 
It makes us ask epistemic questions, namely, what ``reviewing'' is supposed to be.
If a machine can own the verification process and a single human expert can own significance and equity to encounter AI's biases (see, e.g., \cite{zhao2026llmhallucinationswildlargescale}), will additional reviewers be necessary? 
Were the dreaded Reviewers \#2 and 3 simply a means of designing a system that could not originally be automated?

Humans are poor at distinguishing AI-generated work from human texts~\cite{boutadjine2025humanvs}; other fields show similar discrepancies ~\cite{fiedler2025dohumans}.  
The anonymity in reviews, coupled with any intent to deceive and use AI, reveals a standing threat to the assumed trust in reviewing systems.
How the dominoes fall is decided then by accountability, namely, whether expertise is matched, credited, and accounted for -- or replaced by attrition that no one notices until trust fades completely.

\subsection{Whose interests does all of this serve?}

Each of the previous questions comes into focus here, as the future we decide to build together is created not by what is possible but by the goals and actions of many different groups.
Authors of research works want to be accepted quickly, reviewers want an acceptable burden and credit for reviewing, editors and chairs want quality and throughput, venues want prestige and volume, and readers want to trust works they read (see also ~\cite{Schneider2025ofhouse}).
GenAI lowers the brute-force cost of each of these groups simultaneously.

If every role automates the annoying aspects of their position, the system collectively moves more quickly but learns less.
A system in which AI-written papers are reviewed by AI reviewers is locally rational but is wholesale empty.
Provenance and calibration are inert unless accountability is rewarded honestly.
We argue for a transformation: an alignment of incentives to reward meaningful scientific work on all sides.

\section{A Vision for Shaping the Transformation}

We see three grand challenges between the present and any trustworthy use of disruptive technologies such as GenAI in scholarship.
None of these three can be ``solved'' (or even mitigated) by a single actor or even a few; total engagement is required.

\begin{tcolorbox}[title=Three Grand Challenges:,
title filled=false,
colback=violet!5!white,
colframe=violet!40!black,
]
\begin{enumerate}[leftmargin=*]
    \item \textit{Calibrated trust at scale}.
There are no shared reliability benchmarks for GenAI when applied to research artifacts, or even for each individual task in the course of research.
In other words, there is no standard way to determine how often a tool can verify claims, or hallucinates and invents references or misjudges past works' claims.
Without these benchmarks, a flippant ``GenAI helped'' is an assertion rather than evidence, and is not reproducible.
Establishing such benchmarks is on its own a full research agenda we need to tackle.

    \item \textit{Structured provenance within submission systems}, i.e., a machine-readable record of what was tool-assisted and how.
Such functionalities must reside where scholarship is transmitted, namely in submission and other authoring systems.
They require coordination among publishers, professional societies, researchers, tool builders, and research venues. To our knowledge, there is no widely accepted standard for this.

    \item \textit{Integrity when all forces can be synthetic}.
When research artifacts can be easily generated, a trust model based on human authorship is silently broken. The evidence of how humans cannot reliably tell synthetic from human work makes pure detection a lose-lose strategy.
This is not only a technical problem but a \textbf{cultural} one, which is related to our own values.
    \end{enumerate}
\end{tcolorbox}

As a part of our vision, we are proposing directions for research and action that are promising, along with questions to guide them. 
The organizing and foundational principle is a shift from policing GenAI (and other disruptive technologies) to building infrastructure that makes trustworthy scholarship the default. 
We enumerate the next steps for the computing community based on the following philosophical threads, mirroring the previous discussion.


\subsection{Rethinking the Artifact}
It is clear that GenAI models can produce a paper draft directly from underlying data paired with any initial analysis from a human and possibly a verification tool.
In the near term, we see that the research community at large shifts towards having the artifact being \textit{generated from the work} rather than written separately.
In the longer term, the artifact itself can change, such as having the research (meta)data being the primary objects for communicating ideas and meaning, i.e., papers would be only one of many possible ways to present findings.
For example, interactive notebooks, audio/video walk-throughs, and generated dialogue are all currently possible with NotebookLM\footnote{\url{https://notebooklm.google/}} given appropriate data.
Valuing data as scholarly output and improving both provenance and maturity are concrete first steps we can take.


\subsection{Rethinking Review(ers)}

We have many options for near-term courses of action, which are mostly cultural rather than technical.
Open-review models have been tried and tested \cite{ross2017open}, with varying levels of disclosure.
Another option is recognition of reviewing as a first-class scholarly contribution through (among other possibilities) reciprocity, reduced registration and related fees, and citable credit.
An established practice by many conferences is multi-stage review processes, but this, on its own, does not change the status quo.
We believe that the most important near-term action is to offer incentives for \textit{personal, human} feedback: what a reviewer valued (or not), positionality, doubt in either confidence in the results or implications, and so on.
An automated reviewer has no reason to offer this level of feedback.
Indeed, there is large-scale evidence that AI-generated feedback overlaps with that of humans, but skews toward generic statements rather than specific criticisms \cite{liang2024feedback}.

For a longer-term view, we believe that accountability and professional development are the most important prizes.
We envision incorporating reviewing within tenure processes/on curriculum vitae.
Within publication processes, those who do not review responsibly could have their submitted work desk-rejected.
Regarding professional development, authors should have more than one attempt to improve their work. 
None of these requires new technology, but a community that is willing to treat evaluation quality and development on equal footing with output quantity and metrics.

\subsection{Building Calibrated Trust}

If review is to be re-framed as accountable scholarship, then the research community needs shared instruments for determining how reliable a contribution actually is.
One near-term step is to have a venue offer a pilot for a structured, machine-readable provenance schema within the submission system.
Within that, the system records what was AI-assisted and how, and publishes the result so that others can replicate it.
Another near-term step is that the research community can adopt a calibration template, i.e., a short disclosure that any scholarly AI tool completes with a statement on intended use, performance evaluations, and known limitations.
The machine learning community has adopted model cards \cite{mitchell2019model} and datasheets \cite{gebru2021datasheets} for this kind of disclosure.

The long-term view is to build calibrated trust \textit{at scale}.
This can take the form of shared reliability benchmarks for research artifacts and for individual tasks within the research process; these do not exist yet.
Developing these as standard is a research agenda itself, such as defining how to measure a tool's grounding of claims, rate of hallucinated references, tendency to misjudge prior work, and so on.
This cannot be the work of any single actor, but rather coordination among many forces: publishers, professional societies, tool developers, venues, etc., in the same way that the FAIR principles emerged \cite{Wilkinson2016FAIRPrinciples}.
ACM's Artifact Review and Badging \cite{acm2020badging} and similar efforts already ensure that artifacts are documented, consistent, complete, and verified.
This is a natural place where provenance and AI-reliability signals can be merged and layered.
The community should treat such benchmarks and accompanying provenance infrastructure as worth funding.
This is the difference between the two futures, namely that ``GenAI helped'' was an unverifiable assertion in one, and it becomes a calibrated claim in the other.
Here, trust scales with technology rather than eroding away beneath it.

Other efforts to include AI into publication ecosystems are already in development in other related disciplines, such as medical informatics~\cite {pividori2024apublishinginfrastructure}. 
Similarly, journals are starting to adapt towards an AI-augmented editorial infrastructure by relying on epistemological filters, and participatory validation with a human-in-the-loop~\cite{piratova2026curateded}.

\subsection{A Deeper, Cultural Shift}

The more challenging aspect does not lie within the tools we build, but reaffirming what we value as scientists against the pull of quantity, speed, and institutional inertia.
GenAI and similar tools will make us faster either way, but whether it makes us \textit{better} depends on our choices, not those of the models. 
Those choices deserve a communal conversation around broad questions like:
\begin{itemize}[leftmargin=*]
    \item What should count as a scientific contribution? 
    \item How do we want to communicate our ideas in the future? Which audiences, and which modalities?
    \item What principally do we value as scientists? 
    \item What quality do we strive for in academic publications and reviews? 
    \item Which reviewing models can help rebuild trust, and deliver constructive feedback to human researchers? 
    \item How stable would a hypothetical ``idealized'' future be against other disruptive innovations and technologies?
    \item How can we embrace GenAI as technology to become better, more efficient researchers while living up to our ethical values? 
\end{itemize}

We sincerely encourage the entire computing community and all stakeholders to engage in this discussion with us. 
\begin{acks}
Thanks to all our colleagues for the many discussions on this topic, their input, and for sharing thoughts about the recent challenges in computing education research, including but not limited to Jean Blair, Zachary Dodds, Matthew Hicks, Hieke Keuning, Amy Ko, Bendeikt Pfülb, James Prather, Daniel Schiffner, Jan Schneider, and Sandra Schulz. 

The opinions in this work are solely of the authors, and do not necessarily reflect those of the U.S. Army, U.S. Army Research Labs, the U.S. Military Academy, or the Department of Defense.
\end{acks}

\bibliographystyle{ACM-Reference-Format}
\bibliography{sample-base}

\end{document}